\documentclass[preprint,amsmath,amssymb,aps,prc]{revtex4-1}
\usepackage{natbib}
\usepackage{graphicx}
\usepackage{dcolumn}
\usepackage{bm}
\usepackage{gensymb}
\usepackage{tikz}
\usetikzlibrary{plotmarks}
\usepackage{multirow} 
\usepackage{url}
\def\url@leostyle{%
  \@ifundefined{selectfont}{\def\UrlFont{\sf}}{\def\UrlFont{\small\ttfamily}}}
\providecommand*{\unit}[1]{\ensuremath{\mathrm{\,#1}}}
\providecommand*{\unitx}[1]{\ensuremath{\,#1}}
\graphicspath{{figs/}} 

\usepackage{calc}
\usepackage{graphicx}
\usepackage{color}
\begin{document}
\title{The Half-lives of $^{132}$La and $^{135}$La}
\author{E.P. Abel$^1$}
\author{H.K. Clause$^1$}
\author{J. Fonslet$^2$}
\author{R.J. Nickles$^3$}
\author{G.W. Severin$^1$}
\affiliation{$^1$Department of Chemistry and NSCL/FRIB Laboratory, Michigan State University, East Lansing, MI 48824 USA}
\affiliation{$^2$Hevesy Laboratory, Center for Nuclear Technologies, Technical University of Denmark, 4000 Roskilde, Denmark}
\affiliation{$^3$Department of Medical Physics, University of Wisconsin-Madison, Madison WI, 53706 USA}
\date{\today}


\begin{abstract}
The half-lives of $^{135}$La and $^{132}$La were determined via gamma spectroscopy and high-precision ionization chamber measurements. The results are 18.930(6) h for $^{135}$La and 4.59(4) h for $^{132}$La compared to the previously compiled values of 19.5(2) h and 4.8(2) h, respectively. The new results represent an improvement in the precision and accuracy of both values. These lanthanum isotopes comprise a medically interesting system with positron emitter $^{132}$La and Auger electron emitter $^{135}$La forming a matched pair for internal diagnostics and therapeutics. The precise half-lives are necessary for proper evaluation of their value in medicine and for a more representative tabulation of nuclear data.
\end{abstract} 

\maketitle


\begin{center}
\textbf{I. Introduction}
\end{center}

As medical isotopes, $^{132}$La and $^{135}$La can be used together as a matched diagnostic and therapeutic  pair. The advantage of matched pair isotopes in medicine is that the identical chemistry of the isotopes allows an accurate calculation of dosimetry, thereby providing a prediction for the efficacy of internal radionuclide therapy (see for example, the theranostic Terbium isotopes \cite{Mu}). However, inaccuracy in half-life values in either isotope will result in a miscalculation of the dosimetry and systematically lower predictive capability. Therefore, it is important to have accurate and precise nuclear data when performing medical procedures with radionuclides.

In the process of investigating this radionuclide pair, we discovered that not only are the evaluated half-lives relatively imprecise compared to other nuclear data, but they are also inaccurate. Previously measured half-lives for $^{135}$La are shown in Table 1 with the \textit{Nuclear Data Sheets} (NDS) value being a weighted average of the first and second half-lives listed. These values produce the previously accepted half-life of 19.5(2) h for $^{135}$La \cite{NDS135}. Similarly, $^{132}$La was also found to have an inaccurate and imprecise half-life. The accepted half-life of 4.8(2) h for this radionuclide is based on a single value from W.R. Ware and E.O. Wiig in 1960 \cite{NDS132, Wa}.

The present half-life determination uses gamma spectroscopy to elucidate the radionuclide budget in a mixed-species solution and couples the results with high-precision ionization chamber measurements.

\begin{table}[b]
\centering
\caption{\label{table:66gahl} Previous half-life measurements from $^{135}$La decay and the half-life resulting from this work.}
\begin{ruledtabular}
\begin{tabular}{r|ddl}
\textrm{Reference }&\multicolumn{1}{c}{$t_{1/2}$\unit{(h)}}& \multicolumn{1}{c}{$\sigma$\unit{(h)}}\\
\hline
\colrule
\textrm{Morinobu \cite{Mo}, 1965} &19.4&0.1&\\
\textrm{Mitchell \cite{Mi}, 1958 }&19.8&0.2&\\
\textrm{Lavrukhina \cite{La}, 1960} &19.7& - &\\
\textrm{Chubbuck \cite{Ch}, 1948} & 19.5& - &\\
\textrm{Weimer \cite{We}, 1943} &17.5&0.5& \\
\hline 
\textrm{NDS value \cite{NDS135}} & 19.5 & 0.02 &\\
\textrm{This work, 2017 }& 18.930&0.006&\\
\end{tabular} 
\end{ruledtabular}
\end{table}


\begin{center}
\textbf{II. Materials and Methods}
\end{center}

\begin{center}
\textit{A. Source Preparation}
\end{center}

Two separate radiolanthanum sources were prepared for this experiment by the same procedure with slight variations. The values for the first source, “Source 1”, are given in the text with the values for the second source, “Source 2”, appearing in parentheses afterward.

The lanthanum isotopes were created by proton induced reactions on metallic $^{nat}$Ba using a GE-PETtrace cyclotron with nominal incident energy of 16.5 MeV. A 320 mg (640 mg) chunk of barium metal (99.99\% dendritic, Sigma Aldrich), was pressed into an aluminum washer sandwiched between two 50 $\mu$m niobium foils. This was irradiated at 90\degree incidence with a 10 $\mu$A proton beam for 90 min (70 min). After irradiation, the barium and niobium foils were removed from the washer and submerged in 5 mL of 3 M nitric acid (Optima, VWR). The barium dissolved, but the niobium remained intact. A 4 mL (5 mL) fraction of the resulting solution was added to 2 mL (3 mL) of 6 M nitric acid to obtain barium nitrate in a 3 M nitric acid solution. This was passed over 98 mg (103 mg) DGA resin (Eichrom) to trap the lanthanum \cite{Ho}. The resin was washed with 5 mL (10 mL) of 3 N nitric acid, and the lanthanum was eluted with 1 mL of 0.1 M hydrochloric acid (Optima, VWR). Each source elution was split into 200 $\mu$L (500 $\mu$L) fractions: one fraction was taken for gamma spectrum analysis and another was placed in an ionization chamber.

\begin{figure*}
\centering
\scalebox{0.7}{

}
\caption{Decay curves for $^{132}$La and $^{133}$La (top), and $^{135}$La (bottom). The curves were generated from the HPGe data peak-sums at  302, 1909, and 2102 keV from Data Set 2 (top) and 480.5 and 874.5 keV from Data Set 1 and 2 (bottom). The functions resulting from fitting these data points with Equation 2 are plotted as lines for each set of points.}
\end{figure*}
    
\begin{center}
\textit{B. Gamma Spectroscopy}
\end{center}

A fraction of the chemically purified radiolanthanum solution was placed 50 cm (100 cm) from the face of a Canberra C1519 HPGe detector, which was calibrated against $^{133}$Ba, $^{137}$Cs, $^{152}$Eu, $^{241}$Am, and $^{60}$Co standard sources. Counting ensued over the course of the next 90 h (65 h), binning data into 10800 s (900 s) ‘live time’ bins, using an Ortec Aspec 927 ADC and Maestro spectrum software. The ADC deadtime was 32 $\mu$s per analyzed pulse, with initial count rates on the order of $10^{3}$ analyzed pulses per second. After 10 days elapsed, the sample was counted again in the same position to look for long-lived impurities.

Due to overlapping demands on laboratory equipment there was an unavoidable stretch of time during the data collection with Source 2 where $^{72}$As was within the field of view of the detector. Those data were not used in the analysis.

\begin{center}
\textit{C. Ionization Chamber}
\end{center}

A fraction of Source 2 was placed in a refitted Capintec CRC CR-2 re-entrant ionization chamber. The chamber had been previously filled with 10 atm pure argon, and was operated at +300V with voltage control and current measurement by a Keithley 6517a ammeter. The current was logged every 10 s with a LabView controlled ADC for ten days without cessation or removal of the source. During the measurement, the current decayed from 150 nA down to 0.03 nA.  In order to validate the methodology and linearity of the ionization chamber, a $^{64}$Cu source was prepared covering the same range of currents. The $^{64}$Cu source was created by proton bombardment of enriched $^{64}$Ni, and contained small impurities of $^{61}$Cu and $^{61}$Co which were easily distinguished in the decay curve.


\begin{table}[b]
\centering
\caption{\label{table:Activities} Approximate activities for lanthanum isotopes in Source 2 calculated from the 480.5  keV ($^{135}$La),302 keV ($^{133}$La), and 1909 keV ($^{132}$La) peaks.}
\begin{ruledtabular}
\begin{tabular}{c c c}
\multicolumn{1}{c}{Isotope}& \multicolumn{1}{c}{Time after Irradiation\unit{(h)}} & \multicolumn{1}{c}{Activity\unit{(MBq)}}\\
\hline
\colrule
$^{135}$La & 19.9 & 45 \\
$^{133}$La & 2.2 & 13 \\
$^{132}$La & 2.2 & 1.5 \\
\end{tabular} 
\end{ruledtabular}
\end{table}

\begin{center}
\textbf{III. Results}
\end{center}
\begin{center}
\textit{A. Source Characterization}
\end{center}

Gamma spectrum identification of nuclides indicated the presence of $^{131}$La, $^{132}$La, $^{132m}$La, $^{133}$La, $^{135}$La in the earliest measurement, starting 2 hours (55 minutes) after bombardment and 45 minutes (15 minutes) after separation. Undoubtedly, $^{134}$La and $^{136}$La were  co-produced, but had decayed below the detection limit at the beginning of sampling. Also, $^{135m}$Ba was  co-produced in the beam, but was not present in the sample spectra at any time, reflecting the efficiency of the barium removal. Barium isotopes, $^{131}$Ba and $^{133}$Ba, were identified at later time points as daughter products of the lanthanum parents. Within the room there was a background of $^{56}$Co, and the normal natural spectrum from $^{214}$Bi, $^{214}$Pb, and $^{40}$K. As mentioned previously, $^{72}$As was present between hours 18 and 19 during data collection for Source 2 due to other work in the laboratory. No cesium nuclides were identified (as expected due to the low affinity for alkali ions to DGA resin), and the only induced daughter, $^{131}$Cs, was not observed due to the lack of characteristic gamma lines. Also absent was $^{93m}$Mo, which is present in the niobium foil, but was not dissolved into solution or carried through the separation on DGA.The activities of the lanthanum radionuclides at the beginning of the data considered in the determination of the half-lives are given in Table 2.

\begin{center}
\textit{B. Gamma Spectroscopy Results}
\end{center}

For the gamma spectroscopy analysis, data was collected from both Source 1 and 2 to produce Data Set 1 and 2, respectively. Data Set 1 had collection periods of 3 hours and was therefore used only for the determination of the half-life of $^{135}$La. Data Set 2 had much shorter collections periods at 15 minutes each and was used for the half-life measurement for $^{132}$La, $^{133}$La, and $^{135}$La. Characteristic gamma lines were chosen for each of the isotopes: 1909 and 2102 keV for $^{132}$La, 302 keV for $^{133}$La, and 480.5 and 874.5 keV for $^{135}$La. For each of these peaks, a linear baseline correction was calculated by averaging the baseline in surrounding energy bins and was then applied to each of the peak sums. To account for the effect of the dead time on the peak sums, the total counts observed, $O(t_1,t_2)$, in each spectrum between the start time, $t_1$, and the end time, $t_2$, were fit using Equation 1:

\begin{equation}
O(t_1,t_2)=\int_{t_1}^{t_2}R(t)dt=\int_{t_1}^{t_2}B+\sum_{j}\epsilon_jA_je^{-\lambda_jt}dt,
\end{equation}
where $R(t)$ is the total observed count rate. This function is described as a constant background event rate, B, and a series of exponential terms for each nuclide, $j$, present in the sample. Each nuclide also has an associated event detection efficiency, $\epsilon_j$, initial activity, $A_j$, and decay constant, $\lambda_j$. This function was approximated as a background and two exponential terms to capture the effect of the dead time on the peak sums. These two exponential terms corresponded to a shorter and a longer lived species but not to any radionuclide in particular. 
The total observed count rate is then related to the observed peak sums, $S_n(t_1,t_2)$ by Equation 2:

\begin{equation}
S_n(t_1,t_2) = \int_{t_1}^{t_2}(1-rR(t))\epsilon_nA_ie^{-\lambda_it}dt,
\end{equation}
for each peak, $n$ and each radionuclide, $i$. Also, the dead time of 32 $\mu$s per event analyzed is given by $r$. The data for each peak, along with the fitted peak-sum function, are plotted in Figure 1. The half-lives resulting from these fits are given in Table 3.

This method of measuring half-lives was validated by measuring the half-life of $^{133}$La which has a precise and accurate accepted value. This half-life was found to be 3.89(3) h in this work which agrees well with the accepted value of 3.912(8) h \cite{NDS133}.

\begin{table}[b!]
\centering
\caption{\label{table:GammaResults} Measured half-lives for $^{135}$La, $^{132}$La, and $^{133}$La from gamma spectroscopy. }
\begin{ruledtabular}
\begin{tabular}{r|ccc}
\textrm{Data Set}&\multicolumn{1}{c}{Peak Energy\unit{(keV)}}& \multicolumn{1}{c}{$t_{1/2}$\unit{(h)}} &\multicolumn{1}{c}{Average $t_{1/2}$\unit{(h)}}\\
\hline
\colrule
1 & 480.5 & 18.86(4)& \multirow{4}{*}{18.85(3)}\\
1 & 874.5 & 18.9(1) &\\
2 & 480.5 & 18.85(5)&\\
2 & 874.5 & 18.8(2) &\\
\hline
2 & 1909  & 4.71(7) & \multirow{2}{*}{4.59(4)} \\
2 & 2102  & 4.52(5) &\\
\hline
2 & 302   & 3.89(3) & 3.89(3)\\
\end{tabular} 
\end{ruledtabular}
\end{table}

\begin{figure}
{\centering
\begingroup
  \makeatletter
  \providecommand\color[2][]{%
    \GenericError{(gnuplot) \space\space\space\@spaces}{%
      Package color not loaded in conjunction with
      terminal option `colourtext'%
    }{See the gnuplot documentation for explanation.%
    }{Either use 'blacktext' in gnuplot or load the package
      color.sty in LaTeX.}%
    \renewcommand\color[2][]{}%
  }%
  \providecommand\includegraphics[2][]{%
    \GenericError{(gnuplot) \space\space\space\@spaces}{%
      Package graphicx or graphics not loaded%
    }{See the gnuplot documentation for explanation.%
    }{The gnuplot epslatex terminal needs graphicx.sty or graphics.sty.}%
    \renewcommand\includegraphics[2][]{}%
  }%
  \providecommand\rotatebox[2]{#2}%
  \@ifundefined{ifGPcolor}{%
    \newif\ifGPcolor
    \GPcolorfalse
  }{}%
  \@ifundefined{ifGPblacktext}{%
    \newif\ifGPblacktext
    \GPblacktexttrue
  }{}%
  \let\gplgaddtomacro\g@addto@macro
  \gdef\gplbacktext{}%
  \gdef\gplfronttext{}%
  \makeatother
  \ifGPblacktext
    \def\colorrgb#1{}%
    \def\colorgray#1{}%
  \else
    \ifGPcolor
      \def\colorrgb#1{\color[rgb]{#1}}%
      \def\colorgray#1{\color[gray]{#1}}%
      \expandafter\def\csname LTw\endcsname{\color{white}}%
      \expandafter\def\csname LTb\endcsname{\color{black}}%
      \expandafter\def\csname LTa\endcsname{\color{black}}%
      \expandafter\def\csname LT0\endcsname{\color[rgb]{1,0,0}}%
      \expandafter\def\csname LT1\endcsname{\color[rgb]{0,1,0}}%
      \expandafter\def\csname LT2\endcsname{\color[rgb]{0,0,1}}%
      \expandafter\def\csname LT3\endcsname{\color[rgb]{1,0,1}}%
      \expandafter\def\csname LT4\endcsname{\color[rgb]{0,1,1}}%
      \expandafter\def\csname LT5\endcsname{\color[rgb]{1,1,0}}%
      \expandafter\def\csname LT6\endcsname{\color[rgb]{0,0,0}}%
      \expandafter\def\csname LT7\endcsname{\color[rgb]{1,0.3,0}}%
      \expandafter\def\csname LT8\endcsname{\color[rgb]{0.5,0.5,0.5}}%
    \else
      \def\colorrgb#1{\color{black}}%
      \def\colorgray#1{\color[gray]{#1}}%
      \expandafter\def\csname LTw\endcsname{\color{white}}%
      \expandafter\def\csname LTb\endcsname{\color{black}}%
      \expandafter\def\csname LTa\endcsname{\color{black}}%
      \expandafter\def\csname LT0\endcsname{\color{black}}%
      \expandafter\def\csname LT1\endcsname{\color{black}}%
      \expandafter\def\csname LT2\endcsname{\color{black}}%
      \expandafter\def\csname LT3\endcsname{\color{black}}%
      \expandafter\def\csname LT4\endcsname{\color{black}}%
      \expandafter\def\csname LT5\endcsname{\color{black}}%
      \expandafter\def\csname LT6\endcsname{\color{black}}%
      \expandafter\def\csname LT7\endcsname{\color{black}}%
      \expandafter\def\csname LT8\endcsname{\color{black}}%
    \fi
  \fi
    \setlength{\unitlength}{0.0500bp}%
    \ifx\gptboxheight\undefined%
      \newlength{\gptboxheight}%
      \newlength{\gptboxwidth}%
      \newsavebox{\gptboxtext}%
    \fi%
    \setlength{\fboxrule}{0.5pt}%
    \setlength{\fboxsep}{1pt}%
\begin{picture}(7200.00,5040.00)%
    \gplgaddtomacro\gplbacktext{%
      \csname LTb\endcsname%
      \put(726,594){\makebox(0,0)[r]{\strut{} {  $-1$}}}%
      \put(726,1227){\makebox(0,0)[r]{\strut{} {  $-0.5$}}}%
      \put(726,1861){\makebox(0,0)[r]{\strut{} {  $0$}}}%
      \put(726,2494){\makebox(0,0)[r]{\strut{}  {  $0.5$}}}%
      \put(726,3128){\makebox(0,0)[r]{\strut{}  {  $1$}}}%
      \put(726,3761){\makebox(0,0)[r]{\strut{}  {  $1.5$}}}%
      \put(726,4395){\makebox(0,0)[r]{\strut{}  {  $2$}}}%
      \put(858,374){\makebox(0,0){\strut{}  {  $0$}}}%
      \put(1484,374){\makebox(0,0){\strut{} {  $20$}}}%
      \put(2110,374){\makebox(0,0){\strut{} {  $40$}}}%
      \put(2735,374){\makebox(0,0){\strut{}  {  $60$}}}%
      \put(3361,374){\makebox(0,0){\strut{} {  $80$}}}%
      \put(3987,374){\makebox(0,0){\strut{} {  $100$}}}%
      \put(4613,374){\makebox(0,0){\strut{} {  $120$}}}%
      \put(5239,374){\makebox(0,0){\strut{} {  $140$}}}%
      \put(5864,374){\makebox(0,0){\strut{} {  $160$}}}%
      \put(6490,374){\makebox(0,0){\strut{} {  $180$}}}%
    }%
    \gplgaddtomacro\gplfronttext{%
      \csname LTb\endcsname%
      \put(220,2684){\rotatebox{-270}{\makebox(0,0){\strut{} {  Log$_{10}$[Current/nA]}}}}%
      \put(3830,154){\makebox(0,0){\strut{} {  Time (hours)}}}%
      \csname LTb\endcsname%
      \put(5816,4602){\makebox(0,0)[r]{\strut{} {  Data}}}%
      \csname LTb\endcsname%
      \put(5816,4382){\makebox(0,0)[r]{\strut{} {  Fit}}}%
    }%
    \gplbacktext
    \put(0,0){\includegraphics{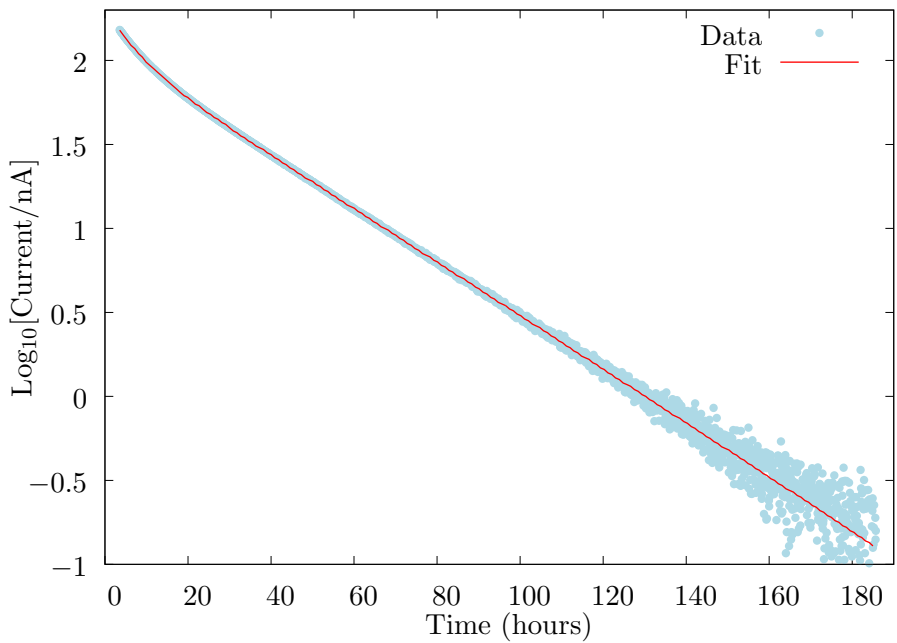}}%
    \gplfronttext
  \end{picture}%
\endgroup

\caption{Plot of current amplitude measurements of the ionization chamber data for $^{135}$La with the resulting fitted function included.}
}
\end{figure}

\begin{center}
\textit{C. Ionization Chamber Results}
\end{center}

Current measurements recorded during the period from 3 hours to 9 days after the separation of Source 2 were used to determine the $^{135}$La half-life. The detector response was fit with a multi-exponential function shown in Equation 3, representing contributions to the signal from $^{135}$La, $^{133}$La, $^{132}$La, $^{132m}$La, $^{131}$La, and $^{131}$Ba:

\begin{equation}
y = \sum_{n}A_ne^{-t\lambda_n}+B.
\end{equation}

This multi-exponential function includes the initial signal, $A_n$, of each nuclide, $n$, at time zero, the respective decay constant of each nuclide, $\lambda_n$, and the time at which the measured current was taken, $t$. A constant background, $B$, was determined after removal of the source, and was used as a subtraction to the data. Only $A_n$ for each nuclide, and the decay constant for $^{135}$La were allowed to float. All other half-lives were fixed during the fit using their respective published values-- with the exception of $^{132}$La, where the measured half-life of 4.59(4) h from the HPGe result in Table 3 was used. 

In order to properly weight the data in the fit, the precision of each current measurement was estimated by examining the intrinsic scatter in the data points. This was done by performing a running average across fifty data points at a time, using twenty-five on either side of a single point of interest and taking the absolute value of the difference between the measured value, $y$, and the running average. These values were treated as residuals, and were smoothed by a running root mean square using one-hundred data points at a time. The smoothed residuals, $\Sigma (x)$, were fit with a function that included a statistical term, represented by the square root of a constant, $C_1$, multiplied by the measured current, $x$, and a constant instrumental term, $C_2$, added in quadrature, as shown in Equation 4:

\begin{equation}
\Sigma(x) = \sqrt{C_1x+C_2^2}.
\end{equation}
This calculated precision, $\Sigma(x)$, was used to weight the data, $y$, for fitting with OriginPro9. The data points and fitting function are shown in Figure 2. The resulting half-life of  $^{135}$La was 18.933(2) h, where the uncertainty only accounts for statistics. 

To determine the presence of any systematics, the methodology was validated using $^{64}$Cu under the same conditions, where $^{61}$Cu and $^{61}$Co were also present. The measured half-life of $^{64}$Cu was 12.6975(5) h compared to the reference value of 12.701(2) h. Assuming that the difference between the evaluated half-life and the ionization chamber measurement results from an imprecision in the experimental methodology, the systematic uncertainty for the measurement was set at 0.28 parts per thousand. Adding this in quadrature to the statistical uncertainty from the $^{135}$La half-life fit results in a final value of 18.933(6) h.

\begin{center}
\textit{D. Final Results}
\end{center}

Through gamma spectroscopy and the ionization chamber measurement, five different half-life measurements were performed for $^{135}$La. These five data points along with the previously accepted half-life are plotted in Figure 3. The values found in this work were averaged using the error of each measurement as respective weights. The final value for the half-life of $^{135}$La was measured to be 18.930(6) h. The error in this case is budgeted as 0.002 h for statistics, and 0.005 h for systematics.  As stated above, the half-life of $^{132}$La was found to be 4.59(4) h, with the precision determined entirely by statistics.

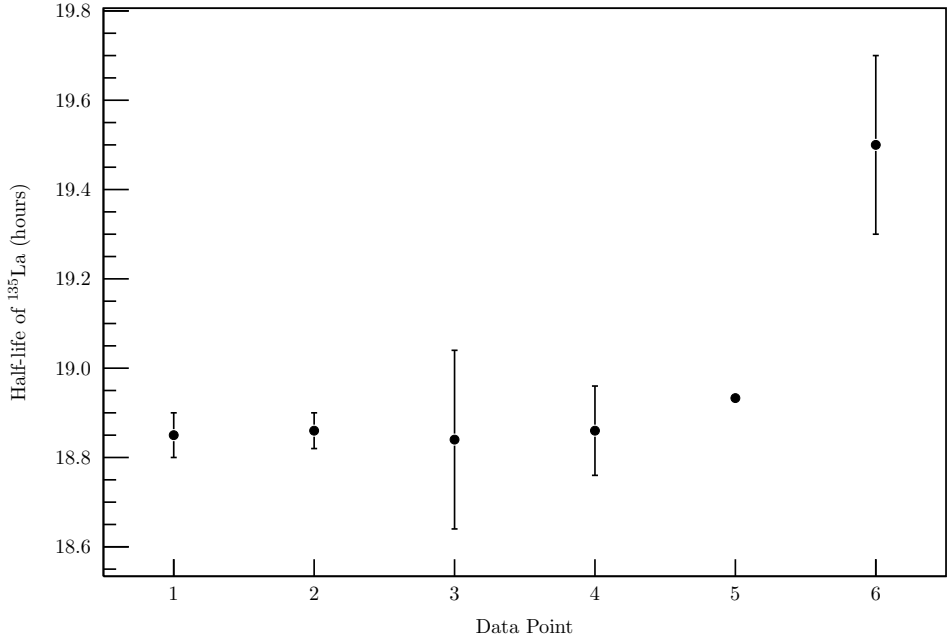
\begin{figure}
{\centering
\scalebox{0.7}{
\begin{tikzpicture}
\pgfdeclareplotmark{cross} {
\pgfpathmoveto{\pgfpoint{-0.3\pgfplotmarksize}{\pgfplotmarksize}}
\pgfpathlineto{\pgfpoint{+0.3\pgfplotmarksize}{\pgfplotmarksize}}
\pgfpathlineto{\pgfpoint{+0.3\pgfplotmarksize}{0.3\pgfplotmarksize}}
\pgfpathlineto{\pgfpoint{+1\pgfplotmarksize}{0.3\pgfplotmarksize}}
\pgfpathlineto{\pgfpoint{+1\pgfplotmarksize}{-0.3\pgfplotmarksize}}
\pgfpathlineto{\pgfpoint{+0.3\pgfplotmarksize}{-0.3\pgfplotmarksize}}
\pgfpathlineto{\pgfpoint{+0.3\pgfplotmarksize}{-1.\pgfplotmarksize}}
\pgfpathlineto{\pgfpoint{-0.3\pgfplotmarksize}{-1.\pgfplotmarksize}}
\pgfpathlineto{\pgfpoint{-0.3\pgfplotmarksize}{-0.3\pgfplotmarksize}}
\pgfpathlineto{\pgfpoint{-1.\pgfplotmarksize}{-0.3\pgfplotmarksize}}
\pgfpathlineto{\pgfpoint{-1.\pgfplotmarksize}{0.3\pgfplotmarksize}}
\pgfpathlineto{\pgfpoint{-0.3\pgfplotmarksize}{0.3\pgfplotmarksize}}
\pgfpathclose
\pgfusepathqstroke
}
\pgfdeclareplotmark{cross*} {
\pgfpathmoveto{\pgfpoint{-0.3\pgfplotmarksize}{\pgfplotmarksize}}
\pgfpathlineto{\pgfpoint{+0.3\pgfplotmarksize}{\pgfplotmarksize}}
\pgfpathlineto{\pgfpoint{+0.3\pgfplotmarksize}{0.3\pgfplotmarksize}}
\pgfpathlineto{\pgfpoint{+1\pgfplotmarksize}{0.3\pgfplotmarksize}}
\pgfpathlineto{\pgfpoint{+1\pgfplotmarksize}{-0.3\pgfplotmarksize}}
\pgfpathlineto{\pgfpoint{+0.3\pgfplotmarksize}{-0.3\pgfplotmarksize}}
\pgfpathlineto{\pgfpoint{+0.3\pgfplotmarksize}{-1.\pgfplotmarksize}}
\pgfpathlineto{\pgfpoint{-0.3\pgfplotmarksize}{-1.\pgfplotmarksize}}
\pgfpathlineto{\pgfpoint{-0.3\pgfplotmarksize}{-0.3\pgfplotmarksize}}
\pgfpathlineto{\pgfpoint{-1.\pgfplotmarksize}{-0.3\pgfplotmarksize}}
\pgfpathlineto{\pgfpoint{-1.\pgfplotmarksize}{0.3\pgfplotmarksize}}
\pgfpathlineto{\pgfpoint{-0.3\pgfplotmarksize}{0.3\pgfplotmarksize}}
\pgfpathclose
\pgfusepathqfillstroke
}
\pgfdeclareplotmark{newstar} {
\pgfpathmoveto{\pgfqpoint{0pt}{\pgfplotmarksize}}
\pgfpathlineto{\pgfqpointpolar{44}{0.5\pgfplotmarksize}}
\pgfpathlineto{\pgfqpointpolar{18}{\pgfplotmarksize}}
\pgfpathlineto{\pgfqpointpolar{-20}{0.5\pgfplotmarksize}}
\pgfpathlineto{\pgfqpointpolar{-54}{\pgfplotmarksize}}
\pgfpathlineto{\pgfqpointpolar{-90}{0.5\pgfplotmarksize}}
\pgfpathlineto{\pgfqpointpolar{234}{\pgfplotmarksize}}
\pgfpathlineto{\pgfqpointpolar{198}{0.5\pgfplotmarksize}}
\pgfpathlineto{\pgfqpointpolar{162}{\pgfplotmarksize}}
\pgfpathlineto{\pgfqpointpolar{134}{0.5\pgfplotmarksize}}
\pgfpathclose
\pgfusepathqstroke
}
\pgfdeclareplotmark{newstar*} {
\pgfpathmoveto{\pgfqpoint{0pt}{\pgfplotmarksize}}
\pgfpathlineto{\pgfqpointpolar{44}{0.5\pgfplotmarksize}}
\pgfpathlineto{\pgfqpointpolar{18}{\pgfplotmarksize}}
\pgfpathlineto{\pgfqpointpolar{-20}{0.5\pgfplotmarksize}}
\pgfpathlineto{\pgfqpointpolar{-54}{\pgfplotmarksize}}
\pgfpathlineto{\pgfqpointpolar{-90}{0.5\pgfplotmarksize}}
\pgfpathlineto{\pgfqpointpolar{234}{\pgfplotmarksize}}
\pgfpathlineto{\pgfqpointpolar{198}{0.5\pgfplotmarksize}}
\pgfpathlineto{\pgfqpointpolar{162}{\pgfplotmarksize}}
\pgfpathlineto{\pgfqpointpolar{134}{0.5\pgfplotmarksize}}
\pgfpathclose
\pgfusepathqfillstroke
}
\definecolor{c}{rgb}{1,1,1};
\draw [color=c, fill=c] (0,0) rectangle (20,13.4957);
\draw [color=c, fill=c] (2,1.34957) rectangle (18,12.1461);
\definecolor{c}{rgb}{0,0,0};
\draw [c,line width=0.9] (2,1.34957) -- (2,12.1461) -- (18,12.1461) -- (18,1.34957) -- (2,1.34957);
\definecolor{c}{rgb}{1,1,1};
\draw [color=c, fill=c] (2,1.34957) rectangle (18,12.1461);
\definecolor{c}{rgb}{0,0,0};
\draw [c,line width=0.9] (2,1.34957) -- (2,12.1461) -- (18,12.1461) -- (18,1.34957) -- (2,1.34957);
\draw [c,line width=0.9] (2,1.34957) -- (18,1.34957);
\draw (10,0.4) node[scale=1, color=c, rotate=0]{Data Point};
\draw [c,line width=0.9] (3.33333,1.67347) -- (3.33333,1.34957);
\draw [c,line width=0.9] (6,1.67347) -- (6,1.34957);
\draw [c,line width=0.9] (8.66667,1.67347) -- (8.66667,1.34957);
\draw [c,line width=0.9] (11.3333,1.67347) -- (11.3333,1.34957);
\draw [c,line width=0.9] (14,1.67347) -- (14,1.34957);
\draw [c,line width=0.9] (16.6667,1.67347) -- (16.6667,1.34957);
\draw [c,line width=0.9] (3.33333,1.67347) -- (3.33333,1.34957);
\draw [c,line width=0.9] (16.6667,1.67347) -- (16.6667,1.34957);
\draw [anchor=base] (3.33333,0.904212) node[scale=1, color=c, rotate=0]{1};
\draw [anchor=base] (6,0.904212) node[scale=1, color=c, rotate=0]{2};
\draw [anchor=base] (8.66667,0.904212) node[scale=1, color=c, rotate=0]{3};
\draw [anchor=base] (11.3333,0.904212) node[scale=1, color=c, rotate=0]{4};
\draw [anchor=base] (14,0.904212) node[scale=1, color=c, rotate=0]{5};
\draw [anchor=base] (16.6667,0.904212) node[scale=1, color=c, rotate=0]{6};
\draw [c,line width=0.9] (2,1.34957) -- (2,12.1461);
\draw (0.4,6.74785) node[scale=1, color=c, rotate=90]{Half-life of $^{135}$La (hours)};
\draw [c,line width=0.9] (2.48,1.90977) -- (2,1.90977);
\draw [c,line width=0.9] (2.24,2.33416) -- (2,2.33416);
\draw [c,line width=0.9] (2.24,2.75856) -- (2,2.75856);
\draw [c,line width=0.9] (2.24,3.18295) -- (2,3.18295);
\draw [c,line width=0.9] (2.48,3.60734) -- (2,3.60734);
\draw [c,line width=0.9] (2.24,4.03173) -- (2,4.03173);
\draw [c,line width=0.9] (2.24,4.45613) -- (2,4.45613);
\draw [c,line width=0.9] (2.24,4.88052) -- (2,4.88052);
\draw [c,line width=0.9] (2.48,5.30491) -- (2,5.30491);
\draw [c,line width=0.9] (2.24,5.72931) -- (2,5.72931);
\draw [c,line width=0.9] (2.24,6.1537) -- (2,6.1537);
\draw [c,line width=0.9] (2.24,6.57809) -- (2,6.57809);
\draw [c,line width=0.9] (2.48,7.00249) -- (2,7.00249);
\draw [c,line width=0.9] (2.24,7.42688) -- (2,7.42688);
\draw [c,line width=0.9] (2.24,7.85127) -- (2,7.85127);
\draw [c,line width=0.9] (2.24,8.27567) -- (2,8.27567);
\draw [c,line width=0.9] (2.48,8.70006) -- (2,8.70006);
\draw [c,line width=0.9] (2.24,9.12445) -- (2,9.12445);
\draw [c,line width=0.9] (2.24,9.54885) -- (2,9.54885);
\draw [c,line width=0.9] (2.24,9.97324) -- (2,9.97324);
\draw [c,line width=0.9] (2.48,10.3976) -- (2,10.3976);
\draw [c,line width=0.9] (2.24,10.822) -- (2,10.822);
\draw [c,line width=0.9] (2.24,11.2464) -- (2,11.2464);
\draw [c,line width=0.9] (2.24,11.6708) -- (2,11.6708);
\draw [c,line width=0.9] (2.48,12.0952) -- (2,12.0952);
\draw [c,line width=0.9] (2.48,1.90977) -- (2,1.90977);
\draw [c,line width=0.9] (2.24,1.48538) -- (2,1.48538);
\draw [c,line width=0.9] (2.48,12.0952) -- (2,12.0952);
\draw [anchor= east] (1.9,1.90977) node[scale=1, color=c, rotate=0]{18.6};
\draw [anchor= east] (1.9,3.60734) node[scale=1, color=c, rotate=0]{18.8};
\draw [anchor= east] (1.9,5.30491) node[scale=1, color=c, rotate=0]{19.0};
\draw [anchor= east] (1.9,7.00249) node[scale=1, color=c, rotate=0]{19.2};
\draw [anchor= east] (1.9,8.70006) node[scale=1, color=c, rotate=0]{19.4};
\draw [anchor= east] (1.9,10.3976) node[scale=1, color=c, rotate=0]{19.6};
\draw [anchor= east] (1.9,12.0952) node[scale=1, color=c, rotate=0]{19.8};
\foreach \P in {(3.33333,4.03173), (6,4.11661), (8.66667,3.94686), (11.3333,4.11661), (14,4.73623), (16.6667,9.54885)}{\draw[mark options={color=c,fill=c},mark size=2.402402pt,mark=*] plot coordinates {\P};}
\draw [c,line width=0.9] (3.33333,4.14635) -- (3.33333,4.45613);
\draw [c,line width=0.9] (3.27603,4.45613) -- (3.39064,4.45613);
\draw [c,line width=0.9] (3.33333,3.91712) -- (3.33333,3.60734);
\draw [c,line width=0.9] (3.27603,3.60734) -- (3.39064,3.60734);
\draw [c,line width=0.9] (6,4.23123) -- (6,4.45613);
\draw [c,line width=0.9] (5.94269,4.45613) -- (6.05731,4.45613);
\draw [c,line width=0.9] (6,4.002) -- (6,3.7771);
\draw [c,line width=0.9] (5.94269,3.7771) -- (6.05731,3.7771);
\draw [c,line width=0.9] (8.66667,4.06147) -- (8.66667,5.64443);
\draw [c,line width=0.9] (8.60936,5.64443) -- (8.72397,5.64443);
\draw [c,line width=0.9] (8.66667,3.83224) -- (8.66667,2.24928);
\draw [c,line width=0.9] (8.60936,2.24928) -- (8.72397,2.24928);
\draw [c,line width=0.9] (11.3333,4.23123) -- (11.3333,4.9654);
\draw [c,line width=0.9] (11.276,4.9654) -- (11.3906,4.9654);
\draw [c,line width=0.9] (11.3333,4.002) -- (11.3333,3.26783);
\draw [c,line width=0.9] (11.276,3.26783) -- (11.3906,3.26783);
\draw [c,line width=0.9] (16.6667,9.66346) -- (16.6667,11.2464);
\draw [c,line width=0.9] (16.6094,11.2464) -- (16.724,11.2464);
\draw [c,line width=0.9] (16.6667,9.43423) -- (16.6667,7.85127);
\draw [c,line width=0.9] (16.6094,7.85127) -- (16.724,7.85127);
\end{tikzpicture}
}

\caption{Plot showing the 5 measurements for the half-life of $^{135}$La in this work and the Nuclear Data Sheet value. Data points 1-5 are the half-lives measured using the 480.5 keV peak in Data Set 2, the 480.5 keV peak in Data Set 1, the 874.5 keV peak in Data Set 2, the 874.5 keV peak in Data Set 1, and the ionization chamber measurement, respectively. Data point 6 is the previously accepted value in NDS. All points, including Data Point 5, are plotted with their respective errors.}
}
\end{figure}


\begin{center}
\textbf{IV. Discussion}
\end{center}

In analyzing the gamma spectroscopy data, careful thought was given to what section of data should be included in the final half-life value. For all of the peaks that were considered, data within the first hour and fifteen minutes was discarded due to a very high count rate in the sample. This count rate would have interfered with fitting the dead time correction and could have introduced many short lived contaminant peaks in the peaks of interest (e.g. $^{132m}$La).

For $^{135}$La, data points starting at 18.92 hours after irradiation for Data Set 1 and 17 hours for Data Set 2 were considered in the half-life values. This data was chosen due to contaminant gamma rays in the 480.5 keV and 874.5 keV peaks for $^{135}$La: 479.5 keV from $^{132}$La and 481.5 keV and 874.8 keV from $^{133}$La. Since $^{132}$La and $^{133}$La had few observed counts by hour 18, these contaminating gamma rays, which already have a low intensity, had an insignificant contribution to either peak. In Data Set 1, one single data point for the 480.5 keV peak (from 23 to 26 hours after irradiation) appears to be an outlier, but there is no known cause. The half-life for $^{135}$La was found for the 480.5 keV peak with and without this supposed outlier, resulting in half-lives of 18.85(4) and 18.83(3) hours, respectively. Clearly, this outlier did not significantly affect the half-life as the two values are the same within their error, and as a result, we included this data point in the value presented in Table 3.

The final time-point used for finding the half-lives of the short-lived lanthanum isotopes was also carefully decided. Since $^{132}$La and $^{133}$La have half-lives on the order of a few hours, only data through 18 hours after the start of the data collection were used in the half-life. At this point, other work in the lab created an increase in the total count rate in the detector, so this was chosen as the cut off for the half-life data for $^{132}$La and $^{133}$La. After this point, very few counts were observed in the peaks for these isotopes showing that this was also a natural stopping point.


\begin{center}
\textbf{V. Conclusion}
\end{center}

Using two different radiolanthanum sources and data from gamma spectroscopy and ionization chamber measurements, the half-lives of $^{132}$La and $^{135}$La were measured. The previously accepted values for these half-lives were found to be both inaccurate and imprecise. In this work, the half-life of $^{132}$La was found to be 4.59(4) hours and the half-life of $^{135}$La was measured as 18.930(6) hours. As these two lanthanum isotopes are medically relevant, these corrected values will aid in calculating the dose that will result from administering these isotopes to patients. In addition, these corrected values provide more accurate and precise data for the nuclear community to use in future work.

\begin{center}
\textbf{Acknowledgements}
\end{center}
We would like to thank Todd Barnhart, Paul Ellison, Stephen Graves, and H{\'e}ctor Valdovinos from the University of Wisconsin Cyclotron Gang for assistance in producing sources and collecting data. This research was supported by The Hevesy Laboratory at the Technical University of Denmark; Michigan State University;  and the National Superconducting Cyclotron Laboratory NSF-1565546.


\begin{thebibliography}{15}

\bibitem{Mu}{C. M{\"u}ller, M. Bunka, S. Haller et.al, }{J. Nucl. Med.} \textbf{55}, 1658 (2014).

\bibitem{Mo}
{S.Morinobu, T.Hirose, and K.Hisatake, } {Nucl.Phys.} \textbf{61}, 613 (1965).


\bibitem{Mi}{A.C.G.Mitchell, C.B.Creager, and C.W.Kocher, } {Phys.Rev.} \textbf{111}, 1343 (1958).

\bibitem{La}{A.K.Lavrukhina, G.M.Kolesov, and Tan Syao–en, }{Columbia Tech. Transl.} \textbf{24}, 1117 (1961).

\bibitem{Ch}{J.B.Chubbuck and I.Perlman, } {Phys.Rev.} 74, 982 (1948).

\bibitem{We}{K.E.Weimer, M.L.Pool, and J.D.Kurbatov, } {Phys.Rev.} 63, 67 (1943).

\bibitem{NDS135}{B. Singh, A.A. Rodionov, and Y.L. Khazov, }{Nuclear Data Sheets} \textbf{109}, 517 (2008).

\bibitem{NDS132}{Yu. Khazov, A.A. Rodionov, B. Singh, and S. Sakharov, }{Nuclear Data Sheets} \textbf{104}, 497 (2005).

\bibitem{Wa}{A.R.Ware and E.O.Wiig, }{Phys. Rev} \textbf{117}, 191 (1960).

\bibitem{Ho}{E.P. Horwitz, D. R. McAlister, A.H. Bond, and R.E. Barrans Jr., }{Solvent Extr. Ion Exc.} \textbf{23}, 319 (2005).

\bibitem{NDS133}{Yu. Khazov, A. Rodionov, and F.G. Kondev, }{Nuclear Data Sheets} \textbf{112}, 855 (2011).

\end{thebibliography}
\end{document}